\documentclass[12pt]{article}
\usepackage{graphicx}

\newsavebox{\sboxpubnumber}
\newsavebox{\sboxpubdate}
\newcommand{\pubdate}[1]{\begin{lrbox}{\sboxpubdate}{#1}\end{lrbox}}
\newcommand{\pubnumber}[1]{\begin{lrbox}{\sboxpubnumber}{\begin{tabular}{l} #1 \\
				 \usebox{\sboxpubdate}
				 \end{tabular}}
                           \end{lrbox}
                           \pubblock}
\newcommand{\Title}[1]{\begin{center} {\Large #1 } \end{center}}
\newcommand{\Author}[1]{\begin{center}{ \sc #1} \end{center}}
\newcommand{\Address}[1]{\begin{center}{ \it #1} \end{center}}

\newcommand{\pubblock}{\rightline{
			\usebox{\sboxpubnumber}}}
\newenvironment{Abstract}{\begin{quotation}  }{\end{quotation}}
\newenvironment{Presented}{\begin{quotation} \hspace{1.85in}PRESENTED AT\bigskip
      \begin{center}\begin{large}}{\end{large}\end{center}
      \end{quotation}}
\newcommand{\Acknowledgements}{\bigskip  \bigskip \begin{center} \begin{large}
             \bf ACKNOWLEDGEMENTS \end{large}\end{center}}
\def\Montreal{Montr\'eal}  
\def\Quebec{Qu\'ebec}

\newcommand{\eq}[1]{equation~(\ref{#1})}
\newcommand{\eqs}[2]{equations~(\ref{#1}) and~(\ref{#2})}
\newcommand{\eqto}[2]{equations~(\ref{#1}) to~(\ref{#2})}
\newcommand{\fig}[1]{Fig.~(\ref{#1})}

\def\g{\gamma}

\def\m{\mu}
\def\n{\nu}
\def\tn{\tilde \nu}

\def\vp{\varphi}
\def\r{\rho}

\def\l{\lambda}
\def\vp{\varphi}

\def\M{{\mathcal M}}  
\def\N{{\mathcal N}}

\def\mT{{\mathbf T}} 

\def\mS{{\mathbf S}}

\begin{document}

\begin{titlepage}
\pubdate{\today}                    
\pubnumber{MCGILL 01-24 \\ {hep-th/0111055}} 
\vspace{.5in}
\vfill
\Title{BRANE GAS COSMOLOGY AND LOITERING}
\vfill
\Author{Damien A. Easson\footnote{Supported 
in part by NSERC (Canada) and FCAR (\Quebec). \\ 
E-mail: \tt easson@hep.physics.mcgill.ca \rm}}
\Address{Physics Department, McGill University\\
      3600 University Street, \Montreal, \Quebec, H3A 2T8 CANADA}
\vfill
\begin{Abstract}
In Brane Gas Cosmology (BGC) the initial state of the universe is taken to be small, dense
and hot, with all fundamental degrees of freedom near thermal
equilibrium.  This starting point is in close analogy with the 
Standard Big Bang (SBB) model. In the simplest example, the topology of the universe is 
assumed to be toroidal in all nine spatial dimensions
and is filled with a gas of $p$-branes.  The dynamics of winding 
modes allow, at most, three spatial dimensions 
to become large, providing a possible
explanation to the origin of our macroscopic (3+1)-dimensional universe.
Specific solutions are found within the model that exhibit loitering,
i.e. the universe experiences a short phase of slow contraction
during which the Hubble radius grows larger than the physical extent of
the universe.  
This phase is studied
by combining the dilaton gravity background equations of motion
with equations that determine the annihilation of string
winding modes into string loops.
Loitering provides a solution to the brane problem (generalised domain wall problem) in BGC
and the horizon problem of the SBB scenario. 
In BGC the initial singularity problem of the SBB scenario is solved, without 
relying on an inflationary phase due to the presence of the T-duality symmetry in the theory.
\end{Abstract}
\vfill
\begin{Presented}
COSMO-01 \\
Rovaniemi, Finland, \\
August 29 -- September 4, 2001
\end{Presented}
\vfill
\end{titlepage}
\def\thefootnote{\fnsymbol{footnote}}
\setcounter{footnote}{0}

\section{Introduction and Motivation}

One of the most significant dilemmas in string theory is the
dimensionality problem. A consistent formulation of superstring theory
requires the universe to be $(9+1)$-dimensional but 
empirical evidence demonstrates that the universe is $(3+1)$-dimensional.  

The typical resolution to this apparent problem 
is to hypothesise that six of the spatial dimensions are curled 
up on a near Planckian sized manifold, and are therefore difficult to
detect in the low energy world that we live in.  
But if this is the case, the question naturally arises, why
is there a difference in size and structure between our large $3$-dimensional
universe and the $6$-dimensional compact space?  What physical
laws demand that spacetime be split in such an unusual way?  Since we
are assuming superstring theory is the correct theory to describe the
physical universe, the answer to these questions must come from within the
theory itself.

Although the dimensionality problem is a very severe problem from a 
cosmological viewpoint, it is rarely addressed.  For example, 
brane world cosmological scenarios derived from string theories 
typically impose the identification
of our universe with a $3$-brane.~\footnote{For an elementary introduction to brane
world constructions see, e.g.~\cite{Easson:2000mj}.}  All current models fail to explain why
our universe is a $d$-brane of spatial dimension $d=3$, opposed to any other
value of $d$, and furthermore fail to explain why our universe is this particular
$3$-brane opposed to any other $3$-brane which may appear in the theory.  Due
to the current unnatural construction of such models, it seems likely that they will
inevitably require some form of the anthropic principle in order to address the
dimensionality problem.

The dimensionality problem is not unique to string theory, however;
it is an equally challenging problem for cosmology.  A truly complete
cosmological model (if it is possible to obtain such a thing), whether derived from
$M$-theory, quantum gravity or any other theory, should necessarily
explain why we live in $(3+1)$-dimensions.

Because this conundrum is an integral part of both superstring theory and
cosmology, it seems likely that only an amalgamation of the two will
be capable of producing a satisfactory solution.  After all, if one is going to
evolve from a $9$-dimensional space to a $3$-dimensional space, one
is going to require dynamics, and the dynamics of our universe are 
governed by cosmology.

In this talk, we consider an approach to string cosmology which addresses
the dimensionality problem and which is in close
analogy with the usual starting point of standard big-bang cosmology.
This scenario is called ``Brane Gas Cosmology" (BGC)~\cite{Alexander:2000xv,Brandenberger:2001kj}.
In BGC, the universe starts out
small, dense, hot and with all fundamental degrees of freedom near thermal 
equilibrium.~\footnote{Note that mathematically it is not possible for the universe to 
be in thermal equilibrium as the FRW cosmological model does not possess a time-like
Killing vector.  However, it is true that the universe has been very nearly in thermal
equilibrium.  Obviously, the departures from equilibrium make things interesting!}
For simplicity, the background spatial geometry is assumed to be toroidal, and the universe is
filled with a hot gas of $p$-branes, the fundamental objects appearing in
string theories.

The motivations for this cosmological scenario, are the problems of the
standard Big Bang model (such as the presence of an initial singularity),
the problems of string theory (such as the dimensionality problem), and are cosmological.  
By cosmological we mean that we wish to stay in close contact with the standard Big Bang
model, and therefore maintain the initial conditions of a hot, small
and dense universe.  Some other attempts to incorporate
$M$-theory into cosmology, such as the existing formulations of brane world scenarios, are motivated
from particle physics and make little connection with what we know about
the origins of the universe.  They also suffer from the dimensionality
problem mentioned above.  Why do the extra dimensions have the topologies they do?
Why should a $3$-brane be favored over any other
$p$-brane for our universe, and why should we live on one particular $3$-three
brane versus another?

Here we present the simplest (and hence, least realistic) model
of BGC.  The universe is assumed to be toroidal, and the brane gas is constructed
from Type IIA supersting theory.  More realistic models with brane gases constructed
from various branches of the $M$-theory moduli space, and with compactifications on manifolds of
non-trivial homology are considered in~\cite{Easson:2001fy}.

The branes may wrap around the cycles of the torus
(winding modes), they can have a center-of-mass motion along the
cycles (momentum modes) or they may simply fluctuate in the bulk space 
(oscillatory modes).
By symmetry, we assume equal numbers of winding and anti-winding modes.
As the universe expands, the winding modes become heavy and halt the 
expansion~\cite{Tseytlin:1992xk}. Spatial dimensions can only
dynamically decompactify if the winding modes can disappear, and this is
only possible (for string winding modes) in $3+1$ 
dimensions~\cite{Brandenberger:1989aj}. Thus, BGC may provide
an explanation for the observed number of large spatial dimensions.
However, causality demands that at least one winding mode per Hubble volume will
be left behind, leading to the {\it brane problem} for 
BGC~\cite{Alexander:2000xv}, a problem analogous to the domain wall
problem of standard cosmology. 

There is a simple solution to the
brane problem: the winding modes will halt the expansion of the spatial
sections, and lead to a phase of slight contraction 
({\it loitering}~\cite{Sahni:1991ks,Feldman:1993ue})
during which the Hubble radius becomes larger than the spatial
sections and hence all remaining winding modes can annihilate in
the large $3+1$ dimensions. We supplement the equations for the
dilaton gravity background of BGC~\cite{Tseytlin:1992xk,Veneziano:1991ek} 
by equations which describe the annihilation of string winding modes
into string loops. Solutions 
exist in which the winding modes cause the universe to contract for
a short time and enter a loitering phase.
During the phase of contraction the number density of the
remaining winding modes increases and the winding
and anti-winding modes begin to annihilate.  The winding branes appear as solitons (analogous to cosmic strings) 
in the bulk space.  The annihilation of winding and antiwinding modes 
(analogous to cosmic string intersections) leads to
the production of string loops which has the same equation of
state as cold matter~\cite{Brandenberger:2001kj}.

Brane Gas Cosmology is a simple nonsingular model which addresses some of 
the problems
of the SBB scenario, simultaneously providing a dynamic resolution to the 
dimensionality problem
of string theory.  

The organisation of this talk is as follows.  
We begin with a brief review of the Brane Gas model in Section~\ref{BG1}.
Our concrete starting point is presented, followed by a derivation of
the equation of state for a gas of branes and an analysis of the background
dynamics.   This is followed (in Section~\ref{LU}) by a discussion of the 
dilaton gravity equations in the presence of a brane gas, of the 
benefits of loitering, and of attractor solutions. In Section~\ref{LP}, we supplement the
system of equations with equations which describe the annihilation of
string winding modes into string loops, and based on this we provide a detailed
analysis of a loitering solution. 
We use both numerical and analytic methods to study the solutions.  
We conclude, in Section~\ref{CONC}, with a brief summary and a few 
conjectures concerning supersymmetry breaking,
the effective breaking of T-duality, and dilaton mass generation 
in the late universe.
\section{Brane Gas Cosmology}\label{BG1}

Over the past decade it has become clear that fundamental strings are not 
the only fundamental degrees of freedom in string theory. D-branes are also 
part of the spectrum of fundamental states. In the Brane Gas scenario we explore some 
possible effects of D-branes on the early universe.  The original
model of~\cite{Alexander:2000xv} is based on 
two key assumptions: firstly that the initial state of the Universe 
corresponded to a dense, hot gas in which all degrees of freedom were in thermal 
equilibrium, and secondly that the topology of the background space admits one-cycles. 
Note that even in spaces without one-cycles in the compactified 
space there are cases where
only three spatial dimensions can become large~\cite{Easson:2001fy}.

The first main point of the scenario in~\cite{Brandenberger:1989aj} is that T-duality will lead to an equivalence of 
the physics if the radius of the background torus changes (in string units) from $R$ to $1/R$. This corresponds 
to an interchange of momentum and winding modes. Thus, $R$ becoming small is equivalent to $R$ tending to infinity. 
Neither limit corresponds to a singularity for string matter. For example, the temperature $T$ obeys
\begin{equation} \label{tempdual}
T({1 \over R}) \, = \, T(R) \, .
\end{equation}
Thus, in string cosmology the big bang singularity can be avoided. 
The second point suggested in \cite{Brandenberger:1989aj} was that string winding modes 
would prevent more than three spatial dimensions from becoming large. 
String winding modes cannot annihilate in more than 
three spatial dimensions (by
a simple classical dimension counting argument). 
In the context of dilaton cosmology,
a gas of string winding modes (which has an equation of state 
$\tilde{p} = - (1/d) \rho$, where $\tilde{p}$ and $\rho$ denote 
pressure and energy density, respectively, and $d$ is the number of 
spatial dimensions) will lead to a confining
potential in the equation of motion for $\lambda = log(a)$, where $a(t)$ is the scale factor of the universe 
\cite{Tseytlin:1992xk}. Note that this is not the result which is obtained in a pure metric background 
obeying the Einstein equations.
The dynamics of classical strings in higher dimensional expanding backgrounds was studied numerically in 
\cite{Sakellariadou:1996vk}, confirming the conclusions of \cite{Brandenberger:1989aj}.

However, it is now clear that string theory has a much richer set of fundamental degrees of freedom, consisting - 
in addition to fundamental strings - of D-branes \cite{Polchinski:1996na} of various dimensionalities. The five previously known
consistent perturbative string theories are now known to be connected by a web of dualities \cite{Witten:1996im}, 
and are believed to represent different corners of moduli space of a yet unknown theory called $M$-theory. Which branes 
arise in the effective string theory description depends on the particular point in moduli space. 

The question we address in~\cite{Alexander:2000xv} is whether the inclusion of the new fundamental degrees of freedom will 
change the main cosmological implications of string theory suggested in \cite{Brandenberger:1989aj}, namely the 
avoidance of the initial cosmological singularity, 
and the singling out of 3 as the maximal number of large spatial dimensions, in the context of an initial state 
which is assumed to be hot, dense and small, and in a background geometry which admits string winding modes.

Our concrete starting point is 11-dimensional $M$-theory compactified on $\mS^1$ to yield 10-dimensional Type IIA string 
theory. The resulting low energy effective theory is supersymmetrized dilaton gravity. As fundamental states, $M$-theory 
admits the graviton, 2-branes and 5-branes. After compactification, this leads to 0-branes, 1-branes, 2-branes, 4-branes, 
5-branes, 6-branes and 8-branes as the fundamental extended objects of the 10-dimensional theory. The dilaton represents 
the radius of the compactified $\mS^1$. We are in a region of moduli space in which the string coupling constant $g_s$ 
is smaller than 1. 

The details of the compactification will not be discussed here,
however we will briefly mention the origins of the above objects
from the fundamental eleven-dimensional, $M$-theory perspective.
The 0-branes of the IIA theory are the BPS states of 
nonvanishing $p_{10}$.  In $M$-theory these are the states of
the massless graviton multiplet.  The 1-brane of the IIA theory
is the fundamental IIA string which is obtained by wrapping
the $M$-theory supermembrane around the $\mS^1$.  The 2-brane
is just the transverse $M2$-brane.  The 4-branes are wrapped
$M5$-branes.  The 5-brane of the IIA theory is a solution 
carrying magnetic NS-NS charge and is an M5-brane that is
transverse to the eleventh dimension.  The 6-brane field
strength is dual to that of the 0-brane, and is a KK
magnetic monopole.  The 8-brane may be viewed as a source for
the dilaton field \cite{Polchinski:1998rr}.  

We assume that all spatial dimensions are toroidal (radius $R$), and that the universe starts out small, dense, hot, and in thermal equilibrium. Thus, the universe will contain a gas of all branes appearing in the spectrum of the theory. Note that this starting point is in close analogy with the hot big bang picture in standard cosmology, but very different from brane-world scenarios in which the existence of a particular set of branes is postulated from the outset without much justification from the point of view of cosmology. 

There have been several interesting studies of the cosmology of brane gases. Maggiore and Riotto \cite{Maggiore:1999cz} (see also \cite{Riotto:2000kn}) studied the phase diagram of brane gases motivated by $M$-theory as a function of the string coupling constant and of
the Hubble expansion rate (as a measure of space-time curvature) and discovered regions of the phase diagram in which brane gases determine the dynamics, and regions where the effective action is no longer well described by a ten-dimensional supergravity action. Given our assumptions, we are in a region in moduli space in which the ten-dimensional effective description of the physics remains true to curvature scales larger than that given by the string scale. In this paper, we consider the time evolution of the system through phase space starting from some
well-defined initial conditions. We will argue that as a consequence of T-duality, curvature scales where the ten-dimensional description breaks down are never reached. 

In another interesting paper, Park et al. \cite{Park:2000xn} take a starting point very close to our own, a hot dense gas of branes. 
However, they did not consider the winding and oscillatory modes of the branes.

In the following section we will study the equation of state of the brane gases for all values of their spatial dimension $p$. 
We will separately analyze the contributions of winding and non-winding modes (the latter treated perturbatively). 
The results will be used as source terms for the equations of motion of the background dilaton gravity fields, following 
the approach of \cite{Tseytlin:1992xk}. We find that the winding modes of any $p$-brane lead to a confining force which prevents 
the expansion of the spatial dimensions, and that the branes with the largest value of $p$ give the largest contribution to the 
energy of the gas in the phase in which the scale factor is increasing.

In Section 3 we argue that the main 
conclusions of the scenario proposed in \cite{Brandenberger:1989aj} are unchanged: T-duality eliminates 
the cosmological singularity, and winding modes only allow three dimensions of space 
to become large. We point out a potential problem (the {\it brane problem}) of  
cosmologies based on theories which admit branes in their spectrum of fundamental 
states. This problem is similar to the well-known domain wall problem \cite{Zeldovich:1974uw} 
in cosmological models based on quantum field theory. It is pointed out that a phase of 
loitering (see e.g. \cite{Sahni:1991ks}) yields a natural solution of this problem, and 
it is shown that the background equations of motion may well yield a 
loitering stage during the early evolution of the universe. Some limitations of 
this model and avenues for future research are discussed in the final section.

\vskip 0.4cm
\section{Equation of State of Brane Gases}

As mentioned in the Introduction, our starting point is Type IIA string 
theory on a 9-dimensional toroidal background space (with the time direction being infinite), resulting from the compactification 
of $M$-theory on $\mS^1$.  The overall spatial manifold has
topology $\mathbf{\M}^{10}_{IIA} =\mS^1 \times \mT^9$.
The fundamental degrees of freedom are the fields of the bulk background (resulting from the graviton in 
$M$-theory), 1-branes, 2-branes, 4-branes, 5-branes, 6-branes and 8-branes.
\newpage
The low-energy bulk effective action is given by
\begin{equation} \label{bulk}
S_{bulk} \, = \, {1 \over {2 \kappa^2}} \int d^{10}x \sqrt{-G} e^{-2 \phi} \bigl[ R + 4 G^{\mu \nu} \nabla_\mu \phi \nabla_\nu \phi
- {1 \over {12}} H_{\mu \nu \alpha}H^{\mu \nu \alpha} \bigr] \, ,
\end{equation}
where $G$ is the determinant of the background metric $G_{\mu \nu}$, $\phi$ is the dilaton, $H$ denotes the field strength corresponding to the bulk antisymmetric tensor field $B_{\mu \nu}$, and $\kappa$ is determined by the 10-dimensional Newton constant in the usual way.

The total action is the sum of the above bulk action and the action of all branes present. The action of an individual brane with spatial dimension $p$ has the Dirac-Born-Infeld form \cite{Polchinski:1996na}
\begin{equation} \label{brane}
S_p \, = \, T_p \int d^{p + 1} \zeta e^{- \phi} \sqrt{- det(g_{mn} + b_{mn} + 2 \pi \alpha' F_{mn})}
\end{equation}
where $T_p$ is the tension of the brane, $g_{mn}$ is the induced metric on the brane, $b_{mn}$ is the induced antisymmetric tensor field, and $F_{mn}$ the field strength tensor of gauge fields $A_m$ living on the brane. The total action is the sum of the 
bulk action (\ref{bulk}) and the sum of all of the brane actions (\ref{brane}), each coupled as a delta function source (a delta function in the directions transverse to the brane) to the 10-dimensional action. 

The induced metric on the brane $g_{mn}$, with indices $m,n,...$ denoting space-time dimensions parallel to the brane, is determined by the background metric $G_{\mu \nu}$ and by scalar fields $\phi_i$ (not to be confused with the dilaton $\phi$) living on the brane (with indices $i,j,...$ denoting dimensions transverse to the brane) which describe the fluctuations of the brane in the transverse directions:
\begin{equation} \label{indmet}
g_{mn} \, = \, G_{mn} + G_{ij} \partial_m \phi_i \partial_n \phi_j + G_{in} \partial_m \phi_i \, .
\end{equation}
The induced antisymmetric tensor field is
\begin{equation} \label{indten}
b_{mn} \, = \, B_{mn} + B_{ij} \partial_m \phi_i \partial_n \phi_j + B_{i[n} \partial_{m]} \phi_i \, .
\end{equation}
In addition,
\begin{equation}
F_{mn} \, = \, \partial_{[m} A_{n]} \, .
\end{equation}

In the string frame, the fundamental string has tension
\begin{equation}
T_f \, = \, (2 \pi \alpha')^{-1} \, ,
\end{equation}
whereas the brane tensions for various values of $p$ are given by \cite{Polchinski:1996na}
\begin{equation}
T_p \, = \, {{\pi} \over {g_s}} (4 \pi^2 \alpha')^{-(p + 1)/2} \, ,
\end{equation}
where $\alpha' \sim l_{st}^2$ is given by the string length scale $l_{st}$ and $g_s$ is 
the string coupling constant.
Note that all of these branes have positive tension.

In the following, we compute the equation of state of the brane gases for a general value of $p$. For our considerations, the most important modes are the winding modes. If the background space is $\mT^9$, a $p$-brane can wrap around any set of $p$ toroidal directions. The modes corresponding to these winding modes by T-duality are the momentum modes corresponding to center of mass motion of the brane. The next most important modes for our considerations are the modes corresponding to fluctuations of the 
brane in transverse directions. These modes are, in the low-energy limit, described by the brane scalar fields $\phi_i$. In addition, there are bulk matter fields and brane matter fields. 

Since we are mainly interested in the effects of a gas of brane winding modes and transverse fluctuations on the evolution of a spatially homogeneous universe, we will neglect the antisymmetric tensor field $B_{\mu \nu}$. We will use conformal time $\eta$
and take the background metric to be given by
\begin{equation}
G_{\mu \nu} \, = \, a(\eta)^2 diag(-1, 1, ..., 1) \, ,
\end{equation}
where $a(\eta)$ is the cosmological scale factor.

If the transverse fluctuations of the brane are small (in the sense that the first term on the right hand side of (\ref{indmet}) dominates) and the gauge fields on the brane are small, then the brane action can be expanded as follows:
\begin{eqnarray} \label{actexp}
S_{brane} \, &=& \, T_p \int d^{p+1} \zeta a(\eta)^{p + 1} e^{-\phi} \times \nonumber \\
& & e^{{1 \over 2} tr log(1 + \partial_m \phi_i \partial_n \phi_i + a(\eta)^{-2} 2 \pi \alpha' F_{mn})} \nonumber \\
&=& \, T_p \int d^{p + 1} \zeta a(\eta)^{p + 1} e^{- \phi} \times \\
& & (1 + {1 \over 2} (\partial_m \phi_i)^2 - \pi^2 {\alpha'}^2 a^{-4} F_{mn}F^{mn}) \, . \nonumber
\end{eqnarray}
The first term in the parentheses in the last line corresponds to the brane winding modes, the second term to the transverse fluctuations, and the third term to brane matter.  We see that, in the low energy limit, the transverse fluctuations of the 
brane are described by a free scalar field action, and the longitudinal fluctuations are given by a Yang-Mills theory. The induced equation of state has pressure $p \geq 0$.

The above result extends to the case of large brane field and brane position fluctuations. It can be shown \cite{Hashimoto:1997gm} that large gauge field fluctuations on the brane give rise to the same equation of state as momentum modes ($E \sim 1/R$) and are thus
also described by pressure $p \geq 0$. In the high energy limit of closely packed branes, the system of transverse brane fluctuations is described by a strongly interacting scalar field theory \cite{Townsend:1999hi} which also corresponds to pressure $p \geq 0$.

We will now consider a gas of branes and determine the equations of state corresponding to the various modes. The procedure involves taking averages of the contributions of all of the branes to the energy-momentum tensor, analogous to what is usually done
in homogeneous cosmology generated by a gas of particles.

Let us first focus on the winding modes. From (\ref{actexp}) it immediately follows that the winding modes of a $p$-brane give rise to the following equation of state:
\begin{equation} \label{EOSwind}
\tilde{p} \, = \, w_p \rho \,\,\, {\rm with} \,\,\, w_p = - {p \over d}
\end{equation}
where $d$ is the number of spatial dimensions (9 in our case), and where $\tilde{p}$ and $\rho$ stand for the pressure and energy density, respectively.

Since both the fluctuations of the branes and brane matter are given by free scalar fields and gauge fields living on the brane (which can be viewed as particles in the transverse directions extended in brane directions), the corresponding equation of state is that of ``ordinary" matter with
\begin{equation} \label{EOSnw}
\tilde{p} \, = \, w \rho \,\,\, {\rm with} \,\,\, 0 \leq w \leq 1 \, .
\end{equation} 
Thus, in the absence of a scalar field sector living on the brane, the energy will not increase as the spatial dimensions expand, in contrast to the energy in the winding modes which evolves according to (as can again be seen immediately from (\ref{actexp}))
\begin{equation} \label{winden}
E_p(a) \, \sim \, T_p a(\eta)^p \, ,
\end{equation}
where the proportionality constant depends on the number of branes. Note that the winding modes of a fundamental string have the same equation of state as that of the winding modes of a 1-brane, and the oscillatory and momentum modes of the string obey the equation of state (\ref{EOSnw}). 

In the context of a hot, dense initial state, the assumption that the brane
fluctuations are small will eventually break down. Higher order terms in the expansion of the brane action will become important. One interesting effect
of these terms is that they will lead to a decrease in the tension of the branes \cite{Maggiore:1999cz}. This will occur when the typical energy scale of the system approaches the string scale. At that point, the state of the system will be dominated by a gas of branes.

The background equations of motion are \cite{Tseytlin:1992xk,Veneziano:1991ek}
\begin{eqnarray} \label{EOMback1}
- d \dot{\lambda}^2 + \dot{\varphi}^2 \, &=& \, e^{\varphi} E \\
\label{EOMback2}
\ddot{\lambda} - \dot{\varphi} \dot{\lambda} \, &=& \, {1 \over 2} e^{\varphi} P \\
\label{EOMback3}
\ddot{\varphi} - d \dot{\lambda}^2 \, &=& \, {1 \over 2} e^{\varphi} E \, ,
\end{eqnarray}
where $E$ and $P$ denote the total energy and pressure, respectively,
\begin{equation}
\lambda(t) \, = \, log (a(t)) \, ,
\end{equation}
and $\varphi$ is a shifted dilaton field which absorbs the space volume factor
\begin{equation}
\varphi \, = \, 2 \phi - d \lambda \, .
\end{equation}
In our context, the matter sources $E$ and $P$ obtain contributions from all components of the brane gas:
\begin{eqnarray}
E \, &=& \, \sum_p E_p^w + E^{nw} \nonumber \\
P \, &=& \, \sum_p w_p E_p^w + w E^{nw} \, ,
\end{eqnarray}
where the superscripts $w$ and $nw$ stand for the winding modes and the non-winding modes, respectively. The contributions of the non-winding modes of all branes have been combined into one term. The constants $w_p$ and $w$ are given by (\ref{EOSwind}) and (\ref{EOSnw}). Each $E_p^w$ is the sum of the energies of all of the brane windings with fixed $p$.

\vspace{0.4cm}
\section{Brane Gases in the Early Universe}

The first important conclusion of \cite{Brandenberger:1989aj} was that in the approach to
string cosmology based on considering string gases in the early universe, the initial cosmological (Big Bang) singularity can be avoided. The question we will now address is whether this conclusion remains true in the presence of branes with $p > 1$ in the spectrum of fundamental states.

The two crucial facts leading to the conclusions of \cite{Brandenberger:1989aj} were T-duality and the fact that in the micro-canonical ensemble the winding modes lead to positive specific heat, leading to {\it limiting} Hagedorn temperature.~\footnote{The Hagedorn temperature is not reached at finite energy density \cite{Abel:1999rq}.}
Both of these facts extend to systems with branes. First, as is obvious, each brane sector by itself preserves T-duality. Secondly, it was shown in \cite{Abel:1999rq} that if two or more Hagedorn systems thermally interact, and at least one of them (let us say System 1) has limiting
Hagedorn temperature, then at temperatures close to the Hagedorn temperature of System 1, most energy flows into that system, and the joint system therefore also has limiting Hagedorn behavior. Hence, as the universe contracts, the T-duality fixed point $R = 1$ is reached at a temperature $T$ smaller than the Hagedorn
temperature, and as the background space contracts further, the temperature starts to decrease according to (\ref{tempdual}). There is no physical singularity as $R$ approaches 0.

Let us now turn to the dynamical de-compactification mechanism of 3 spatial dimensions suggested in \cite{Brandenberger:1989aj}. We assume that the universe starts out
hot, small and in thermal equilibrium, with all spatial dimensions equal (near the self-dual point $R = 1$). In this case, in addition to the momentum and oscillatory modes, winding modes of all $p$-branes will be excited. By symmetry, it is reasonable to assume that all of the net winding numbers cancel, i.e. that there are an equal number of winding and anti-winding modes.

Let us assume that the universe starts expanding symmetrically in all directions. As $\lambda$ increases, the total energy in the winding modes increases according to (\ref{winden}), the contribution of the modes from the branes with the largest value of $p$ growing fastest. In
exact thermal equilibrium energy would flow from the winding modes into non-winding modes. However, this only can occur if the rate of interactions of the winding modes is larger than the Hubble expansion rate.

Generalising the argument of \cite{Brandenberger:1989aj}, from a classical brane point of view it follows (by considering the probability that the world-volumes of two $p$-branes in space-time intersect) that the winding modes of $p$-branes can interact in at most $2p + 1$ large \footnote{Large compared to the string scale.} spatial dimensions. Thus, in $d = 9$ spatial dimensions, there are no obstacles to the disappearance of $p=8$, $p=6$, $p=5$ and $p=4$ winding modes, whereas the lower dimension brane winding modes will allow a hierarchy of dimensions to become large. Since for volumes large compared to the string volume the energy of the branes with the largest value of $p$ is greatest, the 2-branes will have an important effect first. They will only allow 5 spatial dimensions to become large. 
Within this distinguished $\mT^5$, the 1-brane winding modes will only allow a $\mT^3$ subspace to become large.
Thus, it appears that the mechanism proposed in \cite{Brandenberger:1989aj} will also apply if the Hilbert space of states includes fundamental branes with $p > 1$.

To what extent can these classical arguments be trusted? It was shown in \cite{Karliner:1988hd} that the microscopic width of a string increases logarithmically as the energy with which one probes the string.  In our cosmological context, 
we are restricted to energy densities lower than the typical string density, and thus the effective width of the strings is of string scale \cite{Karliner:1988hd}. Similar conclusions will presumably apply to branes of higher dimensionality. However, no definite results are known since a rigorous quantization scheme for higher dimensional branes is lacking. 

The cosmological scenario we have in mind now looks as follows: The universe starts out near the self-dual point as a hot, dense gas of branes, strings and particles. The universe begins to expand in all spatial directions as described by the background equations of motion (\ref{EOMback1} - \ref{EOMback3}). As space expands and cools (and the brane tension therefore increases), the branes will eventually fall out of thermal equilibrium. The branes with the largest value of $p$ will do this first. Space can 
only expand further if the winding modes can annihilate.
This follows immediately from the background equations of motion (\ref{EOMback2}). If the equation of state is dominated by winding modes (which it would be if the universe where to continue expanding), then (with the help of (\ref{EOSwind}) and (\ref{winden})) it follows that the right hand side of that equation acts as if it comes from a confining potential
\begin{equation} \label{pot}
V_{eff}(\lambda) \, = \, \beta_p e^{\varphi} e^{p \lambda} \,
\end{equation}
where $\beta_p$ is a positive constant which depends on the brane tension.

The unwinding of $p$-branes poses no problem for $p=8$, $p=6$, $p=5$ and $p=4$.~\footnote{For a discussion of the microphysics of brane winding mode annihilation see e.g. \cite{Sen:1999mg} and references therein.} The corresponding brane winding modes will disappear first. However, the $p = 2$ branes will then only allow 5 spatial dimensions to expand further (which 5 is determined by thermal fluctuations). 
In this distinguished $\mT^5$, the one-branes and fundamental strings will then only allow a $\mT^3$ subspace to expand.
Thus, there will be a hierarchy of sizes of compact dimensions. In particular, there will be 2 extra spatial dimensions which are larger than the remaining ones.  
A careful counting of dimensions leads to the resulting manifold
\begin{equation}\label{manif}
\mathbf{\M}^{10}_{IIA} = \mS^1 \times \mT^4 \times \mT^2 \times \mT^3
\,,
\end{equation}
where the $\mS^1$ comes from the original compactification of $M$-theory
and the hierarchy of tori are generated by the self-annihilation of $p=2$ and $p=1$ branes
as described above.

From \eq{manif} it appears that the universe may have undergone
a phase during which physics was described by an effective six-dimensional
theory.  It is tempting to draw a relation between this
theory and the scenario of~\cite{Arkani-Hamed:1998rs,Antoniadis:1998ig}
however, since the only scale in the theory is the string scale it
seems unlikely that the extra dimensions are large enough
to solve the hierarchy problem.  This will be studied in a future
publication.

Note that even when winding mode annihilation is possible by dimension counting, causality imposes an obstruction. There will be at least 1 winding mode per Hubble volume remaining (see e.g. \cite{Kibble:1976sj,Brandenberger:1994by}). In our four-dimensional space-time, all branes with $p \geq 2$ will look like domain walls. This leads to the well known domain wall problem \cite{Zeldovich:1974uw} for cosmology since one wall per Hubble volume today will overclose the universe if the tension of the brane is larger than the electroweak scale. Due to their large tension at low temperatures, even one-branes will overclose the universe.

There are two ways to overcome this domain wall problem. The first is to invoke cosmic inflation \cite{Guth:1981zm} at some stage after the branes have fallen out of equilibrium but before they come to dominate the energy of the universe. The scenario would be as follows: initially, the winding modes dominate the energy density and determine the dynamics of space-time. Once they fall out of equilibrium, most will annihilate and the energy in the brane winding modes will become subdominant. The spatial dimensions
in which unwinding occurs will expand. It is at this stage that the ordinary field degrees of 
freedom in the theory must lead to inflation, before the remnant winding modes 
(one per Hubble volume) become dominant.
 
In our context, however, there is another and possibly more appealing 
alternative - loitering \cite{Sahni:1991ks}. If at some stage in the universe the 
Hubble radius  
becomes larger than the spatial extent of the universe, there is no causal obstruction for 
all winding modes to annihilate and disappear.~\footnote{More precisely, the ``causal horizon'', meaning the 
distance which light can travel during the time when $H^{-1}$ is larger than the spatial extent.} In the context of ``standard" cosmology 
(fields as matter sources coupled to classical general relativity) it is very hard 
to obtain loitering. However, in the context of dilaton cosmology a brief phase of 
loitering appears rather naturally as a consequence of the confining potential 
(\ref{pot}) due to the winding modes. To see this, let us return to the background 
equations (\ref{EOMback1} - \ref{EOMback3}). The phase space of solutions was 
discussed for general values of $p$ in \cite{Tseytlin:1992xk}, and a numerical plot of the full phase space is provided in Fig.~1. 

\section{A Loitering Universe}\label{LU}

Within the context of dilaton gravity,
cosmological solutions which exhibit a loitering phase appear rather
naturally due to the presence of winding modes.
The background equations of motion
(\eqto{EOMback1}{EOMback3}) simplify by letting  
$l = {\dot \lambda}$ and $f = {\dot \varphi}$:
\begin{eqnarray}
\label{ldot}
\dot l \, &=& \, {p l^2 \over 2} + lf - \frac{pf^2}{2d} \,,\\
\label{fdot}
\dot f \, &=& \, {d l^2 \over 2} + {f^2 \over 2} \, .
\end{eqnarray}
Notice that for positive energy density $E$, \eq{EOMback1} implies  that $\dot\vp$ will never
change sign.  We are interested in studying the initial conditions with
$\dot\vp < 0$.  If $\dot\vp \not< 0$ the boosting effect of the dilaton on $\l$ will
invalidate the adiabatic approximation used in the derivation of the
EOM.  Furthermore, growing $\vp$ together with expanding $\l$ implies the
growth of the effective coupling $\exp{\phi}$ in contradiction with a weak coupling
assumption~\cite{Tseytlin:1992xk}.  We will therefore consider solutions to the background EOM with initial
conditions corresponding to an expanding universe $\dot \l>0$ with $\dot\vp < 0$.  

For a fixed value of $p$, the phase space of solutions is two-dimensional and is spanned by $l = {\dot \lambda}$ and $f = {\dot \varphi}$. If we start in the energetically allowed (positive $E$) part 
\begin{equation}\label{enal}
\vert l \vert \, < \, {1 \over {\sqrt{d}}} \vert f \vert 
\end{equation}
of the upper left quadrant of phase space with $l > 0$ and $f < 0$
corresponding to expanding solutions with ${\dot \phi} < 0$, then the
solutions are driven towards $l = 0$ at a finite value of $f$ (see
Fig.~1). More precisely, there are three special lines in phase space with \cite{Tseytlin:1992xk}
\begin{equation}
{{\dot l} \over {\dot f}} \, = \, {l \over f} \, ,
\end{equation}
which correspond to straight line trajectories through the origin. They are
\begin{equation}
{l \over f} \, = \, \pm {1 \over {\sqrt{d}}} \, , \, {p \over d} \, .
\end{equation}
Solutions in the energetically allowed part of the upper left quadrant are repelled by the special line 
$l/f = - 1/\sqrt{d}$ and approach $l = 0$. For $l \rightarrow 0$, both $\dot{l}$ and $\dot{f}$ remain finite
\begin{equation}
{\dot l} \, \simeq \, - {p \over {2d}} f^2 \, , \, {\dot f} \, \simeq \, {{f^2} \over 2} \, .
\end{equation}
Hence, the trajectories cross the $l = 0$ axis at some time $t_1$. This means that the expansion of space stops and the universe begins to contract. The crossing time $t_1$ is the first candidate for a loitering point.

Note that the dynamics of the initial collapsing phase are very different from the time reverse of the initial expanding phase. In fact, as the special line $l/f = p/d$ is approached, ${\dot l}$ changes sign again, and the trajectory approaches the static 
solution $(l, f) = (0, 0)$ - also implying a fixed value for the dilaton. At late times, the time evolution along the above mentioned special line corresponds to contraction with a Hubble rate whose absolute value is decreasing,
\begin{equation}
H(t) \, = - {1 \over {\vert H^{-1}(t_0) \vert + \beta (t - t_0)}} \, ,
\end{equation}
(where $t_0$ is some starting time along the special line) with 
\begin{equation}
\beta =  {p \over 2} + {d \over {2p}} \, .
\end{equation}
Thus, the evolution slows down and loitering is reached, even if the time
evolution at $t_1$ is too rapid for loitering to occur then. Note that
these considerations assume that the winding modes are not
decaying. If they decay too quickly, this could obviously prevent
loitering. The above provides an accurate description of the early universe, before 
winding modes have self-annihilated.  We will examine the case of loop production and the
late time evolution of the universe in Section~\ref{LP}.
\begin{figure}\label{phase1.eps}
\centering
\includegraphics[angle=270,width=4in]{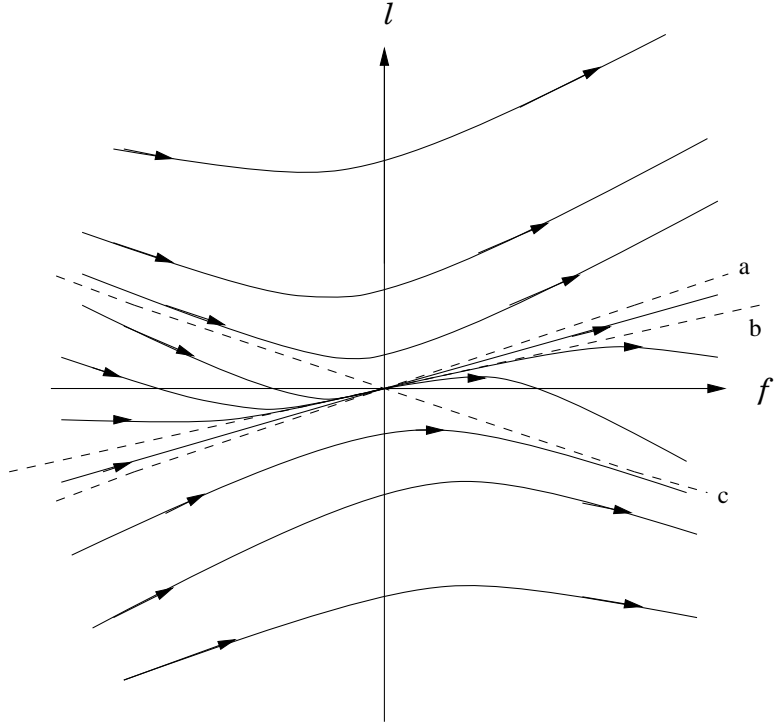}
\caption {Phase space trajectories of the solutions of the background equations (\ref{EOMback1} - \ref{EOMback3}) for the values $p = 2$ and $d = 9$. The energetically allowed region lies near the $l = 0$ axis between the special lines {\it a} and {\it c}, which are the lines given by $l/f = \pm 1 / \sqrt{d}$. The trajectory followed in the scenario investigated in this paper starts out in the upper left quadrant close to the special line {\it c} (corresponding to an expanding background), crosses the $l = 0$ axis at some finite value of $f$ (at this point entering a contracting phase), and then approaches the loitering point $(l, f) = (0, 0)$ along the phase space line {\it b} which corresponds to $l/f = p/d$.}
\end{figure}
Given our cosmological starting point there is no horizon
problem since space was initially of string size. However, two of the
other problems of standard cosmology which the inflationary scenario
successfully addresses, namely the flatness and the formation of
structure problem, persist in our scenario. Note, in particular, that
it seems necessary to have something like inflation of the large
spatial dimensions in order to produce a universe which is larger than
the known Hubble radius. It is of great interest to explore the
possibility of finding a solution of these problems in the context of
string theory (e.g. along the lines of the string-driven inflationary
models of~\cite{Burgess:2001fx}~-~\cite{Turok:1988pg}).
 
\vspace{0.4cm}

\section{Unwinding and Loop Production}\label{LP}

We now wish to extend the analysis presented in Section~\ref{LU} in order to
study the late time behaviour of the universe, 
i.e. to include the effects of winding mode annihilation and loop production.

Recall that after the winding modes have annihilated, a three-dimensional
subspace
will grow large.  In what follows we will therefore take $d=3$.
The strings in the theory are the last branes to unwind which implies that at late times
we should consider the case of $p=1$.  When the winding strings self-annihilate
they create loops in the $(3+1)$-dimensional universe.

We now set up the equations describing the unwinding and corresponding
loop production. They are analogous to the corresponding equations for
cosmic strings in an expanding universe. First, note that the energy
density $\rho_w$ in winding strings can be expressed in terms of the
string tension $\mu$ and the number ${\tilde \nu}(t)$ of winding
modes per ``Hubble" volume $t^3$ as 
\begin{equation} \label{def1}
\r_w(t) \, = \, \m \tn(t) \, t^{-2} \, .
\end{equation}
Since loops are produced by the intersection of two winding strings,
the rate of loop production is proportional to ${\tilde \nu}^2$:
\begin{equation}
\frac{dn(t)}{dt}\, = \, c {\tn(t)}^2 \, t^{-4} \,, 
\end{equation}
where $n(t)$ is the number density of loops and $c$ is a 
proportionality constant expected to be of the order $1$. The
energy density in the winding modes decreases both due to the
expansion of space and due to the decay into loops:
\begin{equation} \label{energy1}
\frac{d\r_w(t)}{dt} + 2 l \, \r_w(t) 
        \, = \, -c'\m t \frac{dn(t)}{dt} \, = \, -cc'\m{\tn(t)}^2 \,
t^{-3} \,,
\end{equation}
where $c^{'}$ is a constant which relates the mean
radius $R = c^{'} t$ of a string loop to its length $\ell$. 
Without loop production ($c = 0$), the energy density 
$\rho_w$ redshifts corresponding to the equation of state
$p = - {1 \over 3} \rho$. This explains the coefficient of
the Hubble damping term in (\ref{energy1}).~\footnote{For
more on strings in an expanding universe see, e.g.~\cite{Vilenkin:1994}.}
Inserting \eq{def1} into the energy conservation \eq{energy1}, we obtain
an equation for $\tn(t)$:
\begin{equation}\label{eomn}
\frac{d\tn(t)}{dt}\, = \, 2 \tn (t^{-1} - l) - cc't^{-1} \tn^2
\,.
\end{equation}

In addition to $\r_w(t)$, we will also require information about the 
energy density in loops, $\r_l(t)$.
The energy density in loops obeys the conservation equation
\begin{equation}\label{julia2}
\frac{d\r_l(t)}{dt} + 3 l \, \r_l(t) 
        \, = \,  cc'\m{\tn(t)}^2 \, t^{-3}
\,.
\end{equation}
As in \eq{energy1}, the second term on the left hand side of
the equation represents the decrease in the density due to
Hubble expansion, with the coefficient reflecting the equation
of state $p = 0$ of a gas of static loops, and the term on the
right hand side representing the energy transfer from winding
modes to loops. Without loop production, $\rho_l(t)$ would
scale as
\begin{equation}\label{rhol}
\r_l(t) \, = \, g(t) e^{-3(\l(t) - \l_0)} 
\, ,
\end{equation}
with $g(t)$ constant. Here $\l_0 = \l(t_0)$, where $t_0$ is some initial
time, and $g(t)$ is a function which obeys the equation
\begin{equation}\label{eomg}
\frac{dg(t)}{dt} \, = \, cc'\m t^{-3} \tn^2 e^{3(\l(t) - \l_0)}
\,.
\end{equation}

Using the expressions for $\r_w(t)$ and $\r_l(t)$ as sources for the energy
density $E$ and pressure $P$ in \eqto{EOMback1}{EOMback3} we can obtain
background equations analogous to \eqs{ldot}{fdot}:
\begin{eqnarray}
\label{eoml}
\dot l \, &=& lf +\frac{1}{2} l^2 - \frac{1}{6} f^2 + \frac{1}{6} g e^{\vp + 3\l_0}\,,\\
\label{eomf}
\dot f \, &=& \, \frac{1}{2} f^2 + \frac{3}{2} l^2 \, .
\end{eqnarray}
(Recall that $p=1$ and $d=3$ in the above equations.)  

Equations (\ref{eomn}), (\ref{eomg}), (\ref{eoml}) and (\ref{eomf}), along with
the equations $l = \dot\l$ and $f = \dot \vp$ provide six, first-order, differential
equations which fully describe
the universe during the process of loop production.
Note that the $c=0$ case corresponds to no loop production and the background equations
reduce to the previous \eqs{ldot}{fdot}.  These provide us with the
initial conditions required for a numerical analysis.

Figure~2 demonstrates the behaviour of a typical numerical solution to the
EOM having initial conditions in the energetically allowed region of the
phase space and accounting for the effects of loop production.  Recall 
that when no loops are produced (see Fig. 1) the 
solutions of interest cross over the $l=0$ line only once and approach
the origin of the phase space as $t \rightarrow \infty$.  When the decay
of the winding modes is taken into account the solutions are pushed back
over the $l=0$ line as in Fig.~2.
\begin{figure}
\centering
\includegraphics[angle=0,width=3in]{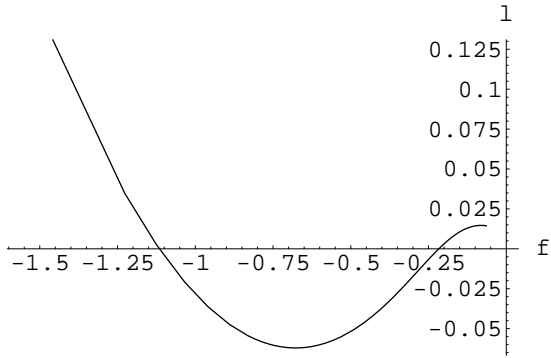}
\caption {A solution of the background equations (\ref{eoml}) and (\ref{eomf}) including
the effects of loop production. 
This depicts a typical solution which starts in the energetically allowed region 
of the phase space. The solution crosses the $l = 0$ axis at some finite value of 
$f$ (at which point the universe enters a contracting  phase), 
and then crosses the
$l=0$ line a second time when the winding modes have fully annihilated.  At this
point the universe begins to expand, is matter dominated and the dilaton is 
assumed to become massive.}
\end{figure}

In more detail, the dynamics of our loitering solution are
depicted in Figures 3, 4, 5 and 6. Figure 3 shows the time evolution
of the Hubble expansion rate $H = l$. Note that since we have set
Newton's constant $G = 1$ in our background equations, time is
measured in Planck time units. 
\begin{figure}\label{tvsl.eps}
\centering
\includegraphics[angle=0,width=3in]{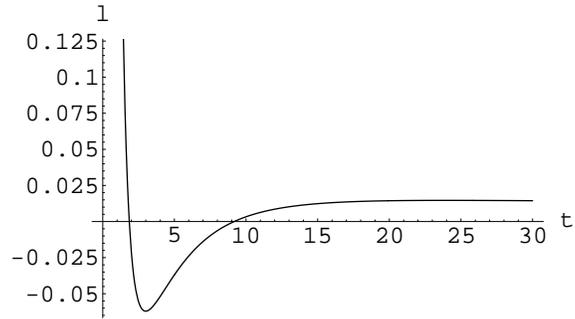}
\caption {The time evolution of $H = l$.  The loitering phase begins
when $l(t)$ crosses the $l =0$ line for the first time and ends when
$l(t)$ crosses back over the $l=0$ line.}
\end{figure}

By comparing the value of
$a(t)$ from Fig. 4, we see
that
\begin{equation} \label{ineq}
H^{-1}(t) \, \gg \, a(t) 
\end{equation}
during the loitering phase. 
\begin{figure}\label{tvsa.eps}
\centering
\includegraphics[angle=0,width=3in]{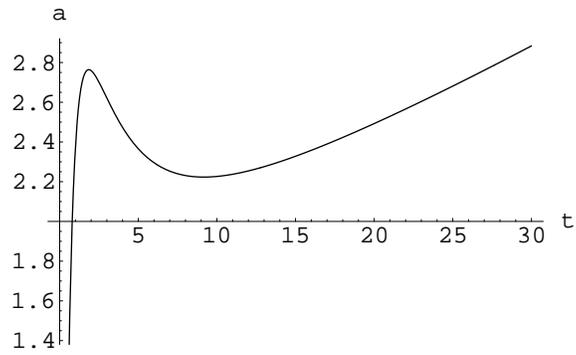}
\caption {The time evolution of the  scale factor $a$.  By comparing
this plot with \fig{tvsl.eps} we see that the loitering phase 
lasts long enough to allow all winding modes to self-annihilate in the 
large three-dimensional universe.}
\end{figure}
Keeping in mind that the initial
spatial size of the tori is Planck scale, it follows immediately
from \eq{ineq} (and from the time duration of the loitering phase)
that loitering lasts sufficiently long to allow causal
communication over the entire spatial section. This is 
reflected in Fig. 5 which shows that the
winding modes completely annihilate by the end of the
loitering phase, after which $g(t)$ tends to a constant (Fig.~6).
\begin{figure}\label{tvsn.eps}
\centering
\includegraphics[angle=0,width=3in]{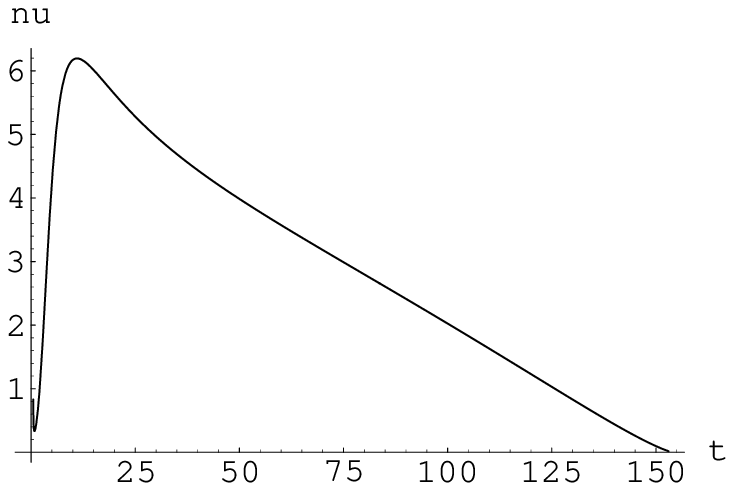}
\caption {Time evolution of $\tilde\n$. Initially, $\tilde\n$
  increases as
the universe contracts.  The winding modes begin to self-annihilate
($\tilde\n$ decreases) and eventually vanish ($\tilde\n
\rightarrow 0$).}
\end{figure}
\begin{figure}\label{tvsg.eps}
\centering
\includegraphics[angle=0,width=3in]{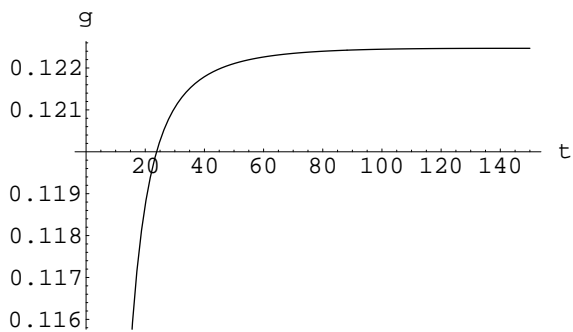}
\caption {Time evolution of $g$.  Note that $g$ goes to a constant
  when
$\tilde\n$ goes to zero.}
\end{figure}

Our modified picture is as follows: the universe begins to expand until 
the winding modes become too massive and force the expansion to stop and contraction to begin.
This corresponds to the solution in Fig.~1 crossing over the $l=0$ axis and
signals the beginning of the loitering phase.  The loitering phase ends when the 
solution is pushed back over the $l=0$ line due to the decay of the
winding modes.  From this point on the universe begins to expand again.

Soon after the solutions cross the $l=0$ line for the second time, our equations become singular.
At the exact moment that the winding modes vanish ($\tn(t) \rightarrow 0$),
all the fields $l$, $f$, $\l$ and $\vp$ in our equations of motion blow up.  
This singularity does not concern us however since it can be eliminated by 
the introduction of a simple
potential $V(\phi)$ used to freeze the dilaton at the moment loop production
is exhausted.  The massless dilaton does not appear in nature and therefore
such a potential is required in any string-theory-motivated cosmological model at late times.
The precise mechanism responsible for dilaton mass generation is unknown,
although it is often suspected that this mechanism will coincide with the breaking
of supersymmetry.  From this analysis we are lead to the conjecture that
dilaton mass generation may coincide with the elimination of winding modes.
We will comment more on this below.

For the time being, let us assume that the dilaton has frozen at the value
$C_\phi = \phi(t_{freeze})$ and therefore $\dot\phi = \ddot\phi = 0$.
We will also assume that this occurs when the winding modes have vanished,
$\tn(t) \rightarrow 0$ and hence the number of 
loops has reached a constant so that $g(t) \rightarrow C_g$.  Now the EOM simplify
greatly.  By fixing the dilaton in \eqs{eoml}{eomf} we can derive an 
equation for the scale factor (after shifting back to the true dilaton $\phi$):
\begin{eqnarray}\label{matdom}
\dot a - C_\g a^{-\frac{1}{2}} = 0
\,,
\end{eqnarray}
where $C_\g$ is a constant given by
\begin{eqnarray}
C_\g = \sqrt{\frac{ C_g }{12}} \, e^{ C_\phi + \frac{3}{2}\l_0}
\,.
\end{eqnarray}
The most general solution to~\eq{matdom} is
\begin{eqnarray}\label{matsol}
a(t) \, = \, \left(\frac{3 C_\g}{2}\right)^{(\frac{2}{3})} \, 
         (t^2 - 2 C t + C^2)^{(\frac{1}{3})}
\,,
\end{eqnarray}
where $C$ is an integration constant.  For the value $C=0$ or for large 
values of $t$ (late times), the scale factor grows as
\begin{equation}\label{mdom}
a(t) \sim t^{\frac{2}{3}}
\,,
\end{equation}
which is exactly the correct behaviour for a matter dominated universe.

Our interpretation is that the winding modes look like solitons in the
$(3+1)$-dimensional universe.  The self-annihilation of these winding modes
corresponds to the creation of matter in the universe and the
scale factor evolves appropriately. In our equations, the loops are
modelled as static. In reality, the loops will oscillate and decay
by emitting (mostly) gravitational radiation, thus producing a radiation
dominated universe.

Let us return to the issue of SUSY breaking and dilaton mass generation.
One thing which appears inevitable within the context of this model 
is the spontaneous ``breaking" of T-duality in the large four-dimensional 
universe.  This is most easily understood once all of the winding 
modes have self-annihilated since it is impossible
to create new ones.  It would cost too much energy for a brane to wrap around
the large dimensions. Thus, the state of the system is not symmetric
under T-duality, and in the absence of string winding modes and for
fixed dilaton, our background equations  
reduce to those of Einstein's General Relativity 
which does not exhibit the $R \leftrightarrow 1/R$ symmetry of string theory.

It is also interesting to note that there seems to be a relation 
between the amount of supersymmetry in a theory and the presence of
T-duality.  Using a specific example in~\cite{Aspinwall:1999ii}, 
Aspinwall and Plesser show that
T-duality can be broken by nonperturbative effects in string coupling. 
Furthermore, a holonomy argument is given to show that T-dualities
should only be expected when large amounts of supersymmetry are present.
It seems likely that the dynamics in the BGC scenario
will cause SUSY to break.  
This result may be in agreement with the possibility of dilaton 
mediated SUSY breaking occurring simultaneously with the breaking of T-duality. 

Considering the above evidence we are inclined to hypothesize about the 
possible relations
between supersymmetry breaking and the breaking of T-duality, as well as
dilaton mass generation and the vanishing of winding modes.

\section{Conclusions and Speculations}\label{CONC}

In this talk we have introduced the model of Brane Gas Cosmology presented in~\cite{Alexander:2000xv}.
The simplest starting point of BGC is $M$-theory compactified on $\mS^1$ which
gives ten-dimensional, Type IIA string theory.  Brane gases constructed from other 
branches of the $M$-theory moduli space are considered in~\cite{Easson:2001fy}.
In the toy model presented here, we assume toroidal topology
in all nine, spatial dimensions.
The initial conditions for the universe
include a small, hot, dense gas of the $p$-branes in the theory.  These
fundamental degrees of freedom are assumed to be in thermal equilibrium.
T-duality ensures that the initial singularity of the SBB model is not present in
this scenario.

We compute the equation of state for the brane gas system and the background
equations of motion.  We study the solutions which initiate in the energetically
allowed region of the phase space.
The universe expands until winding modes
force the expansion to stop and a phase of slow contraction
(loitering) to begin.  Loitering
provides a solution to the brane problem
introduced in~\cite{Alexander:2000xv}.  It also provides solutions to
the horizon problem of the SBB model without relying on an inflationary
phase.

The counting argument of~\cite{Alexander:2000xv} demonstrates that winding modes will allow 
a hierarchy of dimensions in the $\mT^9$ to grow large.  When the string
winding modes self-annihilate we are left with a large $\mT^3$  
subspace, simultaneously
explaining the origin of our $(3+1)$-dimensional universe and solving
the dimensionality problem of string theory.  

Branes wrapped around the cycles of the torus appear as solitonic objects 
in the early universe. They are
topological defects (domain walls for $p \ge 2$).  When the winding
modes and anti-winding modes self-annihilate, matter is 
produced and the universe begins to expand again.
We hypothesize that winding mode annihilation may correspond to dilaton mass
generation.  We also believe there may be a relation between SUSY breaking and
the breaking of T-duality, although we cannot provide any direct evidence for this.
Once winding states have vanished, we cannot
map momentum modes into winding modes via T-duality.  The effective ``breaking''
of T-duality in the large universe requires further study and may be of interest to
finite temperature string theorists.

Particle phenomenology demands compactification on manifolds with
non-trivial holonomy (e.g. Calabi-Yau three-folds)
if the four-dimensional low energy effective theory is to have $\N=1$ 
supersymmetry.  Although we have only examined the trivial toroidal compactification
here, more realistic compactifications are 
considered in~\cite{Easson:2001fy}. 

Brane Gas Cosmology provides an alternative method of incorporating
string and $M$-theory into cosmology to popular ``brane world" scenarios.
In our opinion, BGC has the advantage over such models in that its 
foundations are analogous to those of the Standard Big-Bang model.  
In the BGC model 
the universe starts out small, hot and dense, with no
initial singularity.   

All the current versions of brane
world scenarios embedded in string theories rely on the compactification of the
extra dimensions by hand.  In our opinion this is a considerable problem which
is often overlooked.  A dynamical mechanism in BGC leads naturally to four 
large space-time dimensions. 

\newpage
\Acknowledgements
I would like to thank my collaborators on the 
construction of BGC: Stephon Alexander,
Robert Brandenberger and Dagny Kimberly.  I am grateful to
Cliff Burgess, Jim Cline, Richard Easther, Brian Greene and David Lowe
for many useful discussions during the course of this work.
I would also like to thank the organizers of COSMO-01 and our 
hosts in Roveniemi for providing an interesting and stimulating conference.

\end{document}